\documentclass[twocolumn,apj,twocolappendix]{openjournal}

\usepackage{xcolor}
\usepackage{textgreek}
\usepackage[utf8]{inputenc}
\usepackage[english]{babel}

\usepackage{hyperref}
\hypersetup{
    unicode, 
    colorlinks=true,
    linkcolor=linkcolor,
    citecolor=linkcolor,
    filecolor=linkcolor,
    urlcolor=linkcolor,
}
\usepackage{color,colortbl}
\definecolor{linkcolor}{rgb}{0.0,0.3,0.5}
\usepackage{tensind}
\tensordelimiter{?}
\DeclareGraphicsExtensions{.bmp,.png,.jpg,.pdf}
\usepackage{verbatim}
\usepackage[normalem]{ulem}
\usepackage{orcidlink}
\usepackage{amsmath}
\usepackage{soul}
\usepackage{placeins}

\graphicspath{ {./figs/} }

\begin{document}

\title[Tidal Stripping in Stellar--Dark Matter Systems]{Energy--Space Analysis of Tidal Stripping in Stellar--Dark Matter Systems}

\author{Bradley Arias\orcidlink{0009-0003-5353-3134}}
\email{bradley.arias@berkeley.edu}
\affiliation{Department of Physics, University of California, Berkeley, 110 Sproul Hall, Berkeley, CA 94720, USA}

\author{Nicole E. Drakos\orcidlink{0000-0003-4761-2197}}
\email{ndrakos@hawaii.edu}
\affiliation{Department of Physics and Astronomy, University of Hawaii, Hilo, 200 W Kawili St, Hilo, HI 96720, USA}

\author{James E. Taylor\orcidlink{0000-0002-6639-4183}}
\email{taylor@uwaterloo.ca}
\affiliation{Department of Physics and Astronomy, University of Waterloo, 200 University Ave W, Waterloo, ON N2L 3G1, Canada}
\affiliation{Waterloo Centre for Astrophysics, University of Waterloo,
200 University Avenue West, Waterloo, ON N2L 3G1, Canada}

\begin{abstract}
Observations reveal a striking diversity in dwarf galaxy structures, spanning a wide range of masses, inner density slopes, shapes, and sizes.
Tidal stripping may play a crucial role in shaping the evolution of these galaxies, yet the underlying physical mechanisms remain poorly understood. Using idealized N-body simulations, we investigate the tidal evolution of two-component systems---stellar and dark matter---embedded in a host potential. We find that in terms of energy distributions, both stellar and dark matter particles are stripped identically, regardless of their initial profiles. This surprising result suggests that the energy distribution of stripped stars can provide direct constraints on the underlying dark matter structure. Furthermore, we show that systems with cored dark matter and cuspy stellar profiles naturally evolve into dark matter-deficient (DMD) galaxies, supporting tidal stripping as a viable DMD formation pathway. This energy-space analysis of multi-component systems offers new insights into the dynamical evolution of tidally stripped galaxies.
\end{abstract}


\maketitle

\section{Introduction}

In $\Lambda$CDM, structure formation is predominantly driven by the gravitational collapse of dark matter into gravitationally bound dark matter halos. These halos grow hierarchically, and small merging systems can remain self-bound within a larger halo and are termed subhalos. This dark matter structure provides the scaffolding for structure in the universe, as galaxies form within dark matter halos. While this picture is largely successful at describing our Universe, there are several discrepancies between observations and simulations on dwarf galaxy scales \citep[$M_{*} \lesssim 10^9 M_{\odot}$; for reviews see, e.g.,][]{bullock2017,sales2022}.

Among these discrepancies is the ``galaxy diversity problem"; despite early theoretical predictions, the rotation curves of spiral galaxies have a large range of inner slopes \citep{oman2015}. If the dark matter halos of galaxies are self-similar, as indicated by cosmological simulations \citep[e.g.,][]{navarro1996,navarro1997}, there should be little variation in rotation curves for a fixed maximum circular velocity.  However, many dwarf galaxies have a deficit of mass in the center of the galaxy \citep[originally called the cusp--core problem;][]{moore1994,burkert1995}. A potential solution to this observed diversity in dwarf galaxies is baryonic feedback; the injection of energy from, e.g., stars can form cores in dark matter halos \citep[e.g.,][]{navarro1996,gnedin2002,pontzen2012}. An alternative class of solutions involves modifications to the particle properties of dark matter \citep[e.g., self-interacting dark matter,][]{spergel2000}. 

One particularly interesting class of dwarf galaxies is ultra-diffuse galaxies \citep[UDGs;][]{vandokkum2015}. Although most UDGs are dark matter-dominated \citep[e.g.,][]{vandokkum2016,amorisco2018,sifon2018}, a few notable examples, such as NGC 1052-DF2 and NGC 1052-DF4, have been found to be dark matter-deficient \citep[DMD;][]{vandokkum2018, vandokkum2019}.  Recent work suggests that these systems may represent a distinct class of dwarfs, extending beyond the NGC~1052 group to include other examples such as FCC~224 in the Fornax cluster \citep{buzzo2025}. Although the origins of such DMD galaxies are not well understood, tidal stripping has been proposed as a possible mechanism \citep[e.g.,][]{ogiya2018, carleton2019, nusser2020, ogiya2022, montero2024, contrerassantos2024}. 

In general, idealized simulations---where an isolated N-body subhalo evolves in a fixed background potential---show that, in the absence of baryonic feedback, subhalos follow a characteristic structural evolution. While orbital parameters affect the rate of mass loss, the form of the density profile evolution is largely universal. As first shown by \cite{hayashi2003}, as a dark matter subhalo is stripped, it loses mass primarily on the outside, with some reduction in inner density. Later, \cite{kazantzidis2004} demonstrated that most of the decrease in central density was due to numerical effects, and cuspy subhalos may be expected to survive indefinitely \citep[e.g.,][]{errani2020}. 

In contrast, cored profiles are expected to experience significant decreases in central density until completely disrupted \citep[e.g.,][]{penarrubia2010, drakos2022, errani2023, du2024}. Interestingly, \cite{ogiya2018} demonstrated that dark-matter-deficient galaxies (DMDGs) can be formed in systems with a cored dark-matter profile and a cuspy stellar profile. However, little work has been done studying the dynamics of these multi-component systems. 

Although many models of subhalo evolution are empirical \citep[e.g.,][]{hayashi2003, penarrubia2010, green2019, errani2021}, an intriguing class of models is based on the observation that stripping appears to be simplest in terms of particle energy \citep[e.g.,][]{choi2009, drakos2017, drakos2020, amorisco2021, stucker2021, stucker2023, errani2022, rozier2024}. For example, the energy-truncation model \citep{drakos2017, drakos2020} describes the structural evolution of a dark-matter subhalo by assuming that particles are removed based solely on their initial energy. This model appears universal and reproduces simulation results for both cuspy and cored collisionless systems \citep{drakos2022}. In a similar formulation, the boosted potential framework \citep{stucker2021,stucker2023} redefines the gravitational potential in a moving frame, effectively incorporating tidal effects and providing a natural binding criterion for subhalos. The relationship between these approaches is discussed in \citep{drakos2022}.  These energy-based models offer a simple yet surprisingly accurate model for subhalo evolution under tidal forces. However, these models have so far been limited to single-component systems, and it remains unclear whether their predictive power extends to systems with both stars and dark matter.

This paper aims to examine the tidal evolution of two-component systems in energy space. Specifically, we simulate a collisionless stellar component within a dark matter halo to see how cusps and cores interact and whether these trends are predictable in energy space. In Section~\ref{sec:methods}, we describe the simulations and analysis. In Section~\ref{sec:results}, we show our main results. In Section~\ref{sec:real_gal}, we present a two-component model of NGC 1052-DF2 as a more concrete example. Finally, in Section~\ref{sec:discussion}, we discuss our results.

\section{Simulations} \label{sec:methods}

To investigate the effects of tidal stripping on two-component systems in energy space, we conduct idealized simulations of a satellite orbiting within a fixed background ``host" potential. The satellite is assumed to be spherical and isotropic, an assumption supported by the tendency of stripped systems to evolve towards greater sphericity
\citep{drakos2020}. We neglect contributions from gas, and model the satellite as having two collisionless components---dark matter and stellar---both of which are modeled using a profile we define as the gamma profile. 

\subsection{The gamma profile}

\cite{navarro1996,navarro1997} demonstrated that isolated dark matter halos have a universal profile that is well-approximated by the Navarro-Frenk-White (NFW) profile:
 \begin{equation}
     \rho (r) = \frac{\rho_0}{\frac{r}{r_s}\Big(1+\frac{r}{r_s}\Big)^2}
 \end{equation}
 where $\rho_0$ is a characteristic density and $r_s$ is the scale radius, which corresponds to the point where the logarithmic slope is ${d\ln\rho}/{d\ln r} = -2$.

A more general density profile is given by the ($\alpha$,$\beta$,$\gamma$) double power-law model \citep{jaffe1983, merritt2006},
\begin{equation}
    \rho(r) = \frac{\rho_0}{\Big( \frac{r}{r_s}\Big)^\gamma \Big(1+ \left(\frac{r}{r_s}\right)^\alpha \Big)^{(\beta - \gamma)/\alpha}} \,\,\, .
    \label{eq:powerlaw}
\end{equation}
If we take $\alpha=1$, then $\gamma$ and $\beta$ control the inner and outer slopes of the density profile, respectively.  For $(\alpha, \beta, \gamma) = (1, 3, 1)$, the double-power law is simply the NFW Profile, and for $(\alpha, \beta, \gamma) = (1, 4, 1)$ is the well-known Hernquist Profile \citep{hernquist1990}. 

In this work, we adopt $(\alpha, \beta) = (1, 4)$. Using a value of $\beta = 4$ ensures that the mass profile converges as $r\rightarrow \infty$, which is needed to create a stable isolated system \citep[see, e.g.,][for a discussion]{drakos2017}. Furthermore, the exact value of $\beta$ will not affect the results of this study, as the outer parts of subhalos are quickly stripped to a slope of $-6$ \citep{hayashi2003}.

We then vary $\gamma$ to control the ``cuspiness" of the inner profile. We refer to this as the gamma profile:
\begin{equation}
    \rho_{\gamma}(r) = \frac{\rho_0}{\Big( \frac{r}{r_s}\Big)^\gamma \Big(1+ \frac{r}{r_s} \Big)^{4 - \gamma}} \,\,\ .
    \label{eq:gamma}
\end{equation}
The corresponding enclosed mass profile is given by:
\begin{equation}
    M_{\gamma}(<r) = \frac{4\pi \rho_0 r_s^3}{3-\gamma}\Bigg[\frac{r}{r_s+r} \Bigg]^{3-\gamma} = M_{\rm tot} \Bigg[\frac{r}{r_s+r} \Bigg]^{3-\gamma},
    \label{eq:gamma-mass}
\end{equation}
and the gravitational potential is:
\begin{equation}
    \Phi_\gamma (r) = - \frac{4 \pi G \rho_0 r_s^2}{(3-\gamma)(2-\gamma)}\Bigg[1 - \Bigg(\frac{r}{r + 1}\Bigg)^{2-\gamma}\Bigg] \,\,\, ,
    \label{eq:gamma-pot}
\end{equation}
where $G$ is the gravitational constant.

\subsection{The galaxy model}

We use the gamma profile for both the dark matter ($\rho_{\gamma, \rm dark}$) and stellar components ($\rho_{\gamma, \rm star}$) of the satellite. The total density profile of the satellite is then given by:
\begin{equation}
\rho_{\rm sat}(r) = \rho_{\gamma, \rm dark}(r) + \rho_{\gamma, \rm star}(r) \,\,\, . 
\end{equation}
We consider two different mass and size ratios as summarized in Table~\ref{tab:ICs}. For each satellite configuration, we consider four different combinations of inner slopes, $(\gamma_{\rm star}, \gamma_{\rm dark}) = (0,0)$, $(0,1)$, $(1,0)$ and $(1,1)$, for a total of eight satellite models. Throughout this work, we use the units $G=1$, $r_{\rm unit}=r_s$, and $M_{\rm unit}=M_{\rm sat}$, which correspond to the velocity and time units of $v_{\rm unit}=\sqrt{GM_{\rm sat}/r_s}$ and $ t_{\rm unit} = \sqrt{r_s^3 /(GM_{\rm sat}^3)}$, respectively.

\begin{table}
    \caption{\normalfont Summary of the satellite models, with columns containing (1) name of the satellite, (2) initial dark matter mass: stellar mass, (3) initial scale radius of dark matter component: scale radius of stellar component, (4) initial number of dark matter particles: number of stellar particles.}
    \begin{center}
    \begin{tabular}{c c c c c}
    \hline
     Satellite &
     $M_\mathrm{dark}:M_\mathrm{star}$ &
     $r_\mathrm{dark}:r_\mathrm{star}$ &
     $N_\mathrm{dark}:N_\mathrm{star}$ \\ [0.5ex] 
     \hline\hline
     M10 & $10:1$ & $2:1$ & 909091:90909  \\ 
     M200 & $200:1$ & $10:1$ & 995025:4975  \\ 
     \hline
    \end{tabular}
    \label{tab:ICs}
    \end{center}
\end{table}

\begin{figure}[htp!]
    \includegraphics[]{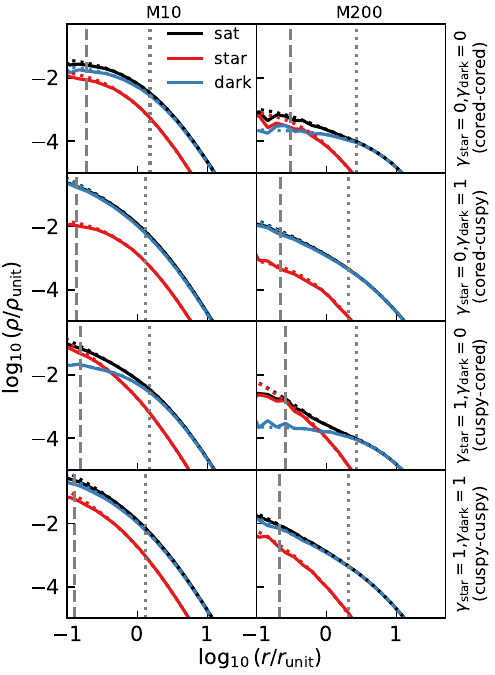}
    \caption{Initial satellite density profiles used in the simulations (dotted lines), given by Equation~\ref{eq:gamma}, with the M10 satellites in the first column and the M200 satellites in the second. The black, red, and blue solid lines indicate the total satellite, the stellar component, and the dark matter component,  respectively. Solid lines show the density profiles of the satellites after they were evolved in isolation for $t=5000 \, t_\mathrm{unit}$.  The gray dotted line shows the relaxation radius $r_\text{relax}$, and the gray dashed line shows the evaporation radius $r_\text{evap}$. The density profiles do not evolve significantly outside the relaxation radius, with a minimal decrease of density outside the evaporation radius.}
    \label{fig:stability}
\end{figure}

\begin{table*}
    \caption{\normalfont Summary of simulation parameters. Columns give (1) name of the satellite simulation (2) mass ratio between dark matter and stellar particles (3) scale radius ratio between the stellar particles and dark matter (4) $\gamma$-value for the stellar component (5) $\gamma$-value for the dark matter component (6) mass ratio between the host and the satellite (7) scale radius ratio between the host and the satellite (8) velocity at apocenter (9) apocenter (10) pericenter,  and (11) orbital period.}
    \centering
    \begin{tabular}{c c c c c c c c c c c c}
     \hline
     Simulation & $M_{\mathrm{dark}}/M_{\mathrm{star}}$  & $r_{\mathrm{s, star}}/r_{\mathrm{s, dark}}$ & $\gamma_\mathrm{star}$ & $\gamma_\mathrm{dark}$ & 
     $M_\mathrm{host}/M_\mathrm{sat}$ & $r_\mathrm{s,host}/r_\mathrm{unit}$ & $v_\mathrm{a}/v_\mathrm{unit}$ & $r_\mathrm{a}/r_\mathrm{unit}$ & $r_\mathrm{p}/r_\mathrm{unit}$ & $t_\mathrm{orb}/t_\mathrm{unit}$ \\[0.5ex] 
     \hline\hline
     M10\_00A & 10 & 2 & 0 & 0 & 200 & 5 & 1.4 & 100 & 63.3 & 240.4 \\ 
     M10\_00B & 10 & 2 & 0 & 0 & 200 & 5 & 0.5 & 200 & 27.0 & 383.1\\ 
     M10\_01A & 10 & 2 & 0 & 1 & 200 & 5 & 1.4 & 100 & 63.3 & 240.4\\
     M10\_01B & 10 & 2 & 0 & 1 & 200 & 5 & 0.5 & 200 & 27.0 & 383.1\\ 
     M10\_10A & 10 & 2 & 1 & 0 & 200 & 5 & 1.4 & 100 & 63.3 & 240.4 \\ 
     M10\_10B & 10 & 2 & 1 & 0 & 200 & 5 & 0.5 & 200 & 27.0 & 383.1\\ 
     M10\_11A & 10 & 2 & 1 & 1 & 200 & 5 & 1.4 & 100 & 63.3 & 240.4 \\ 
     M10\_11B & 10 & 2 & 1 & 1 & 200 & 5 & 0.5 & 200 & 27.0 & 383.1\\ 
     M200\_00A & 200 & 10 & 0 & 0 & 200 & 5 & 1.4 & 100 & 63.3 & 240.4 \\ 
     M200\_00B & 200 & 10 & 0 & 0 & 200 & 5 & 0.5 & 200 & 27.0 & 383.1\\ 
     M200\_01A & 200 & 10 & 0 & 1 & 200 & 5 & 1.4 & 100 & 63.3 & 240.4 \\ 
     M200\_01B & 200 & 10 & 0 & 1 & 200 & 5 & 0.5 & 200 & 27.0 & 383.1\\ 
     M200\_10A & 200 & 10 & 1 & 0 & 200 & 5 & 1.4 & 100 & 63.3 & 240.4 \\ 
     M200\_10B & 200 & 10 & 1 & 0 & 200 & 5 & 0.5 & 200 & 27.0 & 383.1\\ 
     M200\_11A & 200 & 10 & 1 & 1 & 200 & 5 & 1.4 & 100 & 63.3 & 240.4 \\ 
     M200\_11B & 200 & 10 & 1 & 1 & 200 & 5 & 0.5 & 200 & 27.0 & 383.1\\ 
     \hline
    \end{tabular}
    \label{tab:sims}
\end{table*}

We generate initial conditions for $10^6$ equal-mass particles\footnote{Unequal particle masses can cause spurious heating in cosmological simulations, as noted in \cite{ludlow2019, ludlow2023}. This effect is avoided in our simulations by using equal particle masses for dark matter and stars.},
by sampling positions and velocities from their component density profiles and distribution functions in the usual manner \citep[e.g.,][]{kazantzidis2004, drakos2017}. For multi-component systems, the distribution function for each component, $i$, is given by:
\begin{equation}
f_i(\mathcal{E}) =  \frac{1}{\sqrt{8} \pi^2} \left[ \int_{r_{\mathcal{E}}}^\infty\frac{1}{\sqrt{\mathcal{E} - \Psi}} \frac{{\rm d}^2 \rho_i}{{\rm d} \Psi^2} \frac{GM(<r)}{r^2} \rm{d}r \right] \,\,\, ,
\end{equation}
where $\Psi(r)=-\Phi(r)$ is the total potential of the satellite, $\mathcal{E}(r,v)= \Psi(r)-v^2/2$ is the binding energy,  $M(<r)$ is the total mass of the satellite within radius $r$,  and $\rho_i$ is the density profile of component $i$. The lower limit of integration, $r_{\mathcal{E}}$ is defined as the radius at which the binding energy is equal to the potential energy, i.e., $\mathcal{E}~=~\Psi(r_{\mathcal{E}})$.

To verify the stability of the initial conditions, we evolved the satellite models to $t = 5000\,t_\mathrm{unit}$ in isolation (i.e., no background potential) with a softening length of $0.1\,r_\mathrm{unit}$ using the N-body code \textsc{Gadget-2} \citep{springel2005}. Figure~\ref{fig:stability} shows that the M10 satellites remain stable over time, with neither the dark matter nor stellar component exhibiting significant density evolution beyond $r > 0.5\,r_{\rm unit}$. Additionally, we note that the timescale shown here ($5000\,t_\mathrm{unit}$) is much longer than the simulations shown in the next section. 

In Figure~\ref{fig:stability}, we also show the relaxation (vertical dashed lines) and evaporation radius (vertical dotted lines). Inside these radii, numerical collisional effects are expected to become important at simulation time $t$. Specifically, the relaxation timescale, 
\begin{equation}
t_{\rm rel} (r) \approx 0.1 \dfrac{\sqrt{N(<r)}}{\ln N(<r)}\sqrt{\dfrac{r^3}{GM}}, 
\end{equation}
where $N(<r)$ is the number of particles within radius $r$, is the time in which a typical particle's velocity changes by an order of itself. The evaporation timescale, $t_{\rm evap} (r)\approx 136 t_{\rm rel} (r)$, is the time it takes a typical particle to reach its escape speed.

\begin{figure}[htp!]
    \centering
 \includegraphics[width=\columnwidth]{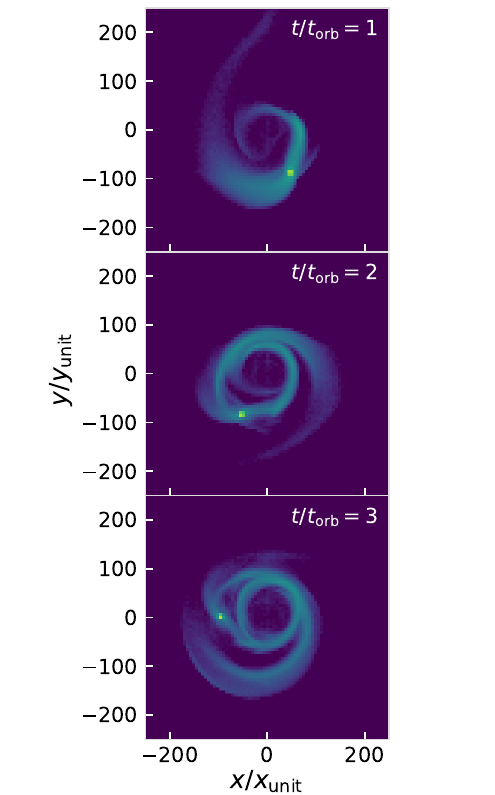}
    \caption{Simulation M10\_10A. Plots show the binned density of all particles (dark matter and stellar) in $x$ and $y$-coordinates for time $t/t_\mathrm{orb} = 1$ (top), $t_\mathrm{orb} = 3$ (middle), and  $t/t_\mathrm{orb} = 5$ (bottom). Particles are removed in streams, and a bound center remains.}
    \label{fig:system_plot}
\end{figure}

\subsection{Orbital parameters}

We evolved these eight satellites within a host potential in two different orbital configurations
termed $A$ and $B$
for a total of 16 simulations, as summarized in Table~\ref{tab:sims}. All simulations were run using a modified version of \textsc{Gadget-2}, with the host potential modeled as an NFW profile with $M_{\rm host}(<50\, r_{\rm unit})=200 M_{r_{\rm unit}}$ and $r_{s, \rm host} = 5 r_{\rm unit}$. To illustrate a typical case, we present the M10\_10A simulation in Figure~\ref{fig:system_plot}.

\subsection{Bound remnants}

To identify the bound satellite remnant, we first locate the satellite’s frame by centering on the highest density peak of the particle distribution, as described in \citep[e.g.,][]{tormen1997, drakos2017}. This is done by initially defining a sphere of radius $R$ that encompasses all particles (only considering particles that were self-bound in the previous time step) located at the center of mass (COM) of the satellite. We then iteratively decrease the radius of the sphere to $0.9\,R$ and recenter on the new COM. The process is completed when the sphere contains fewer than 100 particles. The velocity frame of the satellite is determined in the same manner.

Given this definition of the satellite frame, we can calculate the ``binding" energy for each particle as
\begin{equation}
    \mathcal{E} = -\Phi(r) - \frac{1}{2}v^2 \,\,\, ,
\label{eq:binding_energy}
\end{equation}
where $r$ and $v$ are with respect to the satellite center. Assuming the system is spherical and isotropic, the self-potential of the satellite, $\Phi$, can be approximated as in \cite{drakos2020}:
\begin{equation}
    \Phi(r_i) = -Gm \Bigg( \frac{N(<r_i)}{r_i} + \sum_{j=r_j>r_i}^N \frac{1}{r_j} \Bigg) \,\,\, ,
\end{equation}
where $N$ is the number of particles, $r_i$ is the radius of the $i$th particle and $N(<r_i)$ is the number of particles within $r_i$. 

To determine which particles are self-bound, we calculate the binding energy of all the particles and remove any unbound particles ($\mathcal{E}<0$), then recalculate the potential. This process is repeated until convergence, leaving the particles that are self-bound. We note that this procedure neglects the contribution of the background potential, using only the potential of the bound satellite particles.  In the next section, we explore the evolution of the bound remnant. 

\section{Evolution of the bound remnant} \label{sec:results}

In this section, we present our main findings on mass loss from two-component satellites (Section~\ref{sec:massloss}), the evolution of their density profiles (Section~\ref{sec:density-profile}), and their behavior in energy space (Section~\ref{sec:energy-space}). We find similar trends in all the simulations, but we mainly focus on the M10\_XXA simulations, where the mass loss proceeds more slowly, and the trends are clearer. 

\subsection{Mass loss}\label{sec:massloss}

\begin{figure}[htp!]
    \includegraphics[]{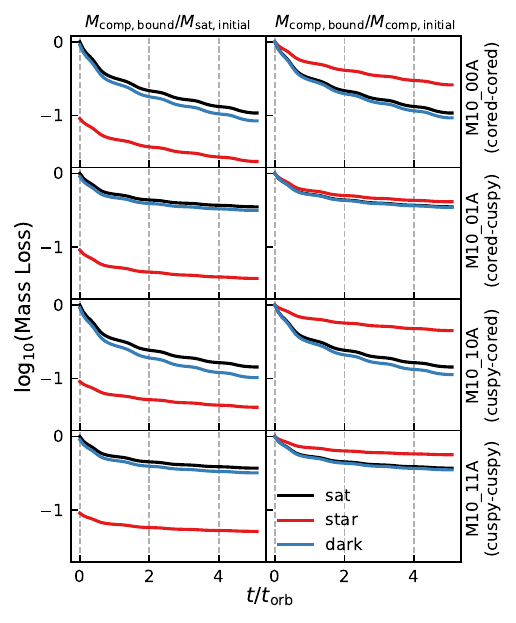}
     \caption{Mass loss of M10 satellites. $M_\mathrm{comp,bound}/M_\mathrm{sat,total}$ is shown on the left column and $M_\mathrm{comp,bound}/M_\mathrm{comp,total}$ on the right, with black lines corresponding to the satellite (comp = sat), red to the stellar component (comp = star), and blue to the dark matter component (comp = dark). In satellites with cuspy components, particles are stripped more slowly, allowing bound remnants to survive longer.}
    \label{fig:massloss}
\end{figure}

Figure~\ref{fig:massloss} shows the bound mass fraction of each component compared to either the total initial satellite mass ($M_\mathrm{comp, bound}/M_\mathrm{sat,total}$) or the initial mass of that component ($M_\mathrm{comp, bound}/M_\mathrm{comp,total}$). From these plots, we see that the satellites lose mass as they orbit, with most mass loss occurring at pericenter, consistent with previous results \citep[e.g.,][]{hayashi2003}. This mass loss is successively smaller after each orbit, with simulations containing cored components being stripped faster than cuspy components as expected \citep{penarrubia2010, drakos2022}. Across the simulations, the dark matter components lose mass more rapidly than the stellar components, relative to their respective initial masses (right column). Moreover, satellites containing cuspy components remain bound longer before complete disruption, as cuspy components effectively ``protect" the other component from stripping.

\begin{figure}[htp!]
    \includegraphics[]{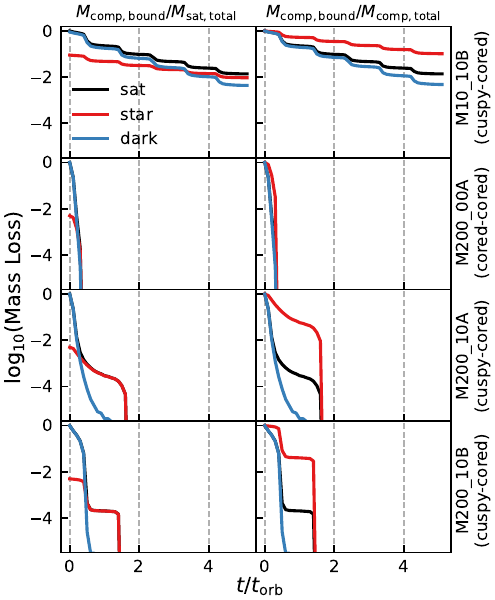}
     \caption{Mass-loss in simulations that result in stellar-dominated systems.  $M_\mathrm{comp,bound}/M_\mathrm{sat,total}$ is shown on the left column and $M_\mathrm{comp,bound}/M_\mathrm{comp,total}$ on the right, with black lines corresponding to the satellite (comp = sat), red to the stellar component (comp = star), and blue to the dark matter component (comp = dark). We find that these stellar-dominated systems typically arise when the distribution of stars is cuspy and the distribution of dark matter is cored.}
    \label{fig:lowdark}
\end{figure}

In certain cases, if the dark matter is cored, it can be stripped more than the stellar component, resulting in a satellite that is lacking in dark matter, as shown in Figure~\ref{fig:lowdark}. Typically, we only found this to occur in cases where the star distribution was cuspy, with one possible exception (M200\_00A). In most of these simulations, the satellite was typically completely disrupted soon after losing its dark matter. However, it is unclear whether this is partially due to numerical effects (see discussion in Section~\ref{sec:discussion}). 

\subsection{Density profile evolution}\label{sec:density-profile}

\begin{figure*}[htp!]
	\includegraphics[]{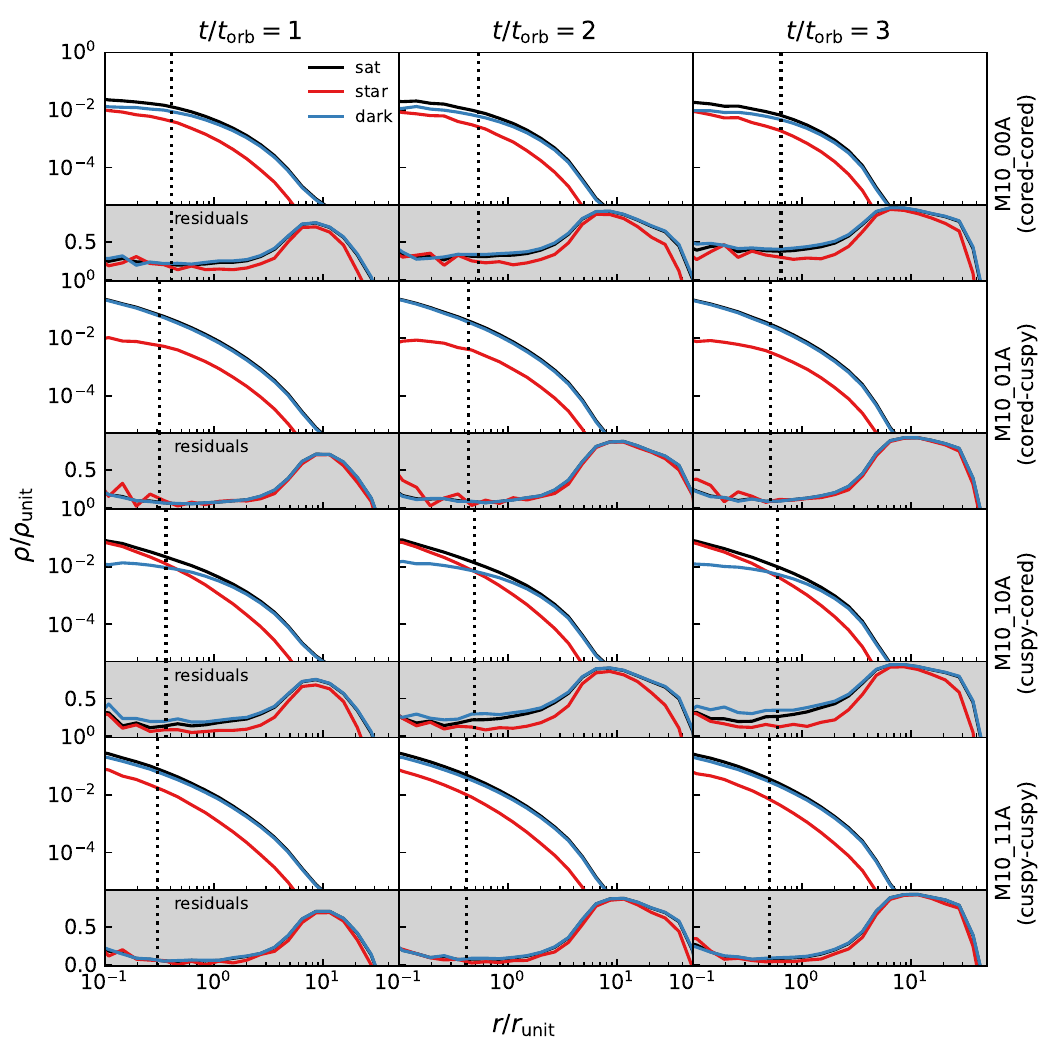}
    \caption{Density profile evolution of M10\_XXA simulations. The solid lines show $\rho(r)$ for stellar particles (red), dark matter particles (blue), and the satellite total (black). The columns of plots correspond to different times (from $t/t_\mathrm{orb}=1$ to $t/t_\mathrm{orb}=3$). The plots show that particles are preferentially stripped from large radii, with some reduction in central density. Simulations with cuspy components lose less of their central densities compared to simulations with cored components. The dotted black lines indicate the relaxation radius at that time.  The residuals (grey boxes), defined as $(\rho_{\text{comp},0}-\rho_{\text{comp},t})/\rho_{\text{comp},0}$, show that the cuspy--cuspy simulation has the smallest reduction in central density, while the cored--cored simulation has the largest reduction.}
    \label{fig:density_M10_R2}
\end{figure*}

Figure~\ref{fig:density_M10_R2} shows the evolution of the satellite density profiles, $\rho(r)$, and residuals $(\rho_{\text{comp},0}-\rho_{\text{comp},t})/\rho_{\text{comp},0}$. As the satellite orbits around the host, it loses mass. Preferentially, particles at larger radii are stripped, but there is also some reduction in central density (ranging from 5--25 percent after the first orbit), particularly for cored--cored profiles \citep[as expected from, e.g.,][]{penarrubia2010, drakos2022}. This difference arises because cored and cuspy components respond differently to tidal mass loss. While cuspy components lose mass primarily from their outskirts, cored components can lose material from a wider range of initial radii \citep[e.g.][]{drakos2022}. Particles at small radii in cuspy systems are typically among the most tightly bound, whereas particles at the same radius in cored systems span a broader range of binding energies. As a result, tidal stripping can remove material from the inner regions of cored components more efficiently, leading to larger fractional reductions in density at fixed radius.

In mixed gamma cases---$(\gamma_{\rm star},\gamma_{\rm dark})=(0, 1)$ and $(\gamma_{\rm star},\gamma_{\rm dark})=(1, 0)$---the cuspy component mitigates the drop in central density of the cored component, although it does not entirely prevent it. 
Finally, we note that at very large radii, the residuals decrease again. This is likely due to material that is temporarily bound but is leaving the system \citep{penarrubia2009, drakos2020}.

\subsection{Energy-space stripping}\label{sec:energy-space}

As suggested in previous studies \cite[e.g.,][]{choi2009, drakos2017}, tidal stripping may be best understood in terms of particle energies. In this section, we follow \cite{choi2009} and describe the angular momentum and energy of each particle in terms of circularity, $\epsilon_{\mathrm circ}$, and binding energy, $\mathcal{E}$, given by Equation~\ref{eq:binding_energy}.

For a particle with radius $r$ from the center of the satellite and velocity $v$ from the rest frame of the satellite, the circularity is defined as:
\begin{equation}
    \epsilon_{\rm circ} = \frac{L}{L_{\rm max}}
\end{equation}
where $L$ is the angular momentum of the particle,
\begin{equation}
    L = |\vec{r} \times \vec{v}|
\end{equation}
and $L_{\rm max}$ is the maximum angular momentum a particle with this energy could have, 
\begin{equation}
    L_{\rm max} = \sqrt{GM(<r_{\mathcal{E}})r_{\mathcal{E}}}
\end{equation}
 where $r_{\mathcal{E}}$ is the radius of a circular orbit with energy $\mathcal{E}$, and $M(<r_{\mathcal{E}})$ is the mass within this radius. This gives a circularity value $\epsilon_{\rm circ}$ that ranges between 0 and 1 for each particle's initial velocity and radius, where $\epsilon_\mathrm{circ} = 1$ is a circular orbit.

\begin{figure}[htp!]
\includegraphics[]{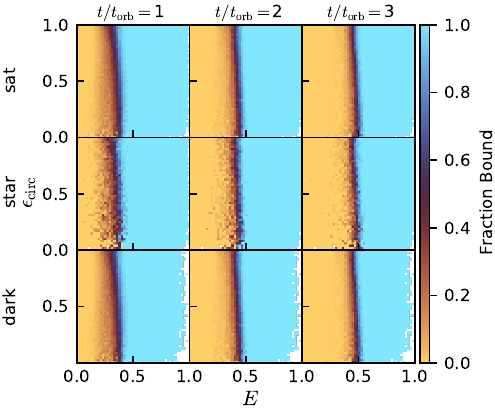}
    \caption{Energy-space stripping for M10\_10A.
    We show a 2D histogram of particles in the initial energy--circularity plane, with normalized energy $E$ on the $x$-axis and circularity $\epsilon_{\rm circ}$ on the y-axis. The color indicates the fraction of initially bound particles remaining. The histograms show all the satellite particles (top row), the stellar component (middle row), and the dark matter component (bottom row). The columns show the stripping at different orbits $t/t_\mathrm{orb}$. Both components are stripped to the same energy value despite the fact that they are non-self-similar.}
    \label{fig:energy_stripping}
\end{figure}

\begin{figure}[htp!]
\includegraphics[]{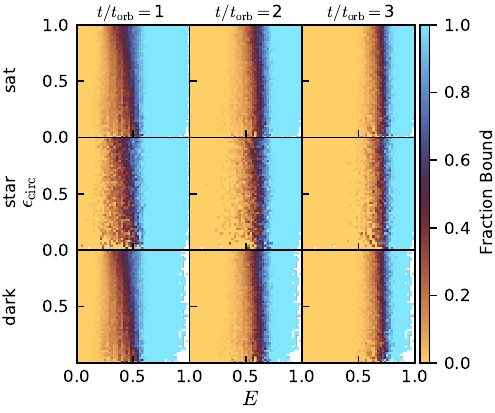}
    \caption{Energy-space stripping for M10\_10B, shown for comparison with M10\_10A in Figure~\ref{fig:energy_stripping}. As in M10\_10A, both components are stripped to the same truncation energy. However, in this case, the truncation is less sharply defined, with a more gradual fall-off in particle fraction at the energy boundary.}
    \label{fig:appendix_energy}
\end{figure}

The binding energies of the particles in the satellite frame, defined by Equation~\ref{eq:binding_energy}, are normalized by using the initial central potential energy of the satellite:
\begin{equation}
    \Phi_0 = -\frac{4\pi G \rho_\mathrm{s,star} r_\mathrm{s,star}^2}{(3-\gamma_\mathrm{star})(2-\gamma_\mathrm{star})} - \frac{4\pi G \rho_\mathrm{s,dark} r_\mathrm{s,dark}^2}{(3-\gamma_\mathrm{dark})(2-\gamma_\mathrm{dark})}
\end{equation}

The normalized binding energy is denoted as 
\begin{equation}
E = \dfrac{\mathcal{E}}{\Psi_0} = -\dfrac{\mathcal{E}}{\Phi_0} 
\end{equation}
where particles with $E \in [0,1]$ are said to be bound. 

We bin the particles by their initial circularity $\epsilon_\mathrm{circ}$ and their initial normalized binding energies $E$. At each snapshot, we remove unbound particles (with $E<0$) and then calculate the fraction of particles remaining in each bin, relative to the initial distribution. This is shown for the M10\_10A simulation in Figure~\ref{fig:energy_stripping}. We see that particles are stripped primarily according to their initial binding energy, with some slight dependence on circularity. Notably, stellar and dark matter particles are stripped identically in this initial energy–angular momentum space, even when their distributions are not self-similar. 

Similarly, in Figure~\ref{fig:appendix_energy}, we show the energy–circularity distribution for the M10\_10B simulation. As with M10\_01A, particles in both components are stripped to similar initial energies, with a slight dependence on orbital circularity, though the transition at the truncation energy is less sharp. As shown in \cite{drakos2020}, this is because the assumptions underlying simple energy-truncation models become less accurate for more eccentric orbits, as individual particles undergo greater mixing in energy and angular momentum during an orbital period. The M200 simulations (not shown) follow the same general pattern, but rapid disruption also leads to noisier and less informative energy-space plots.

\begin{figure*}[htp!]
\includegraphics[scale=0.85]{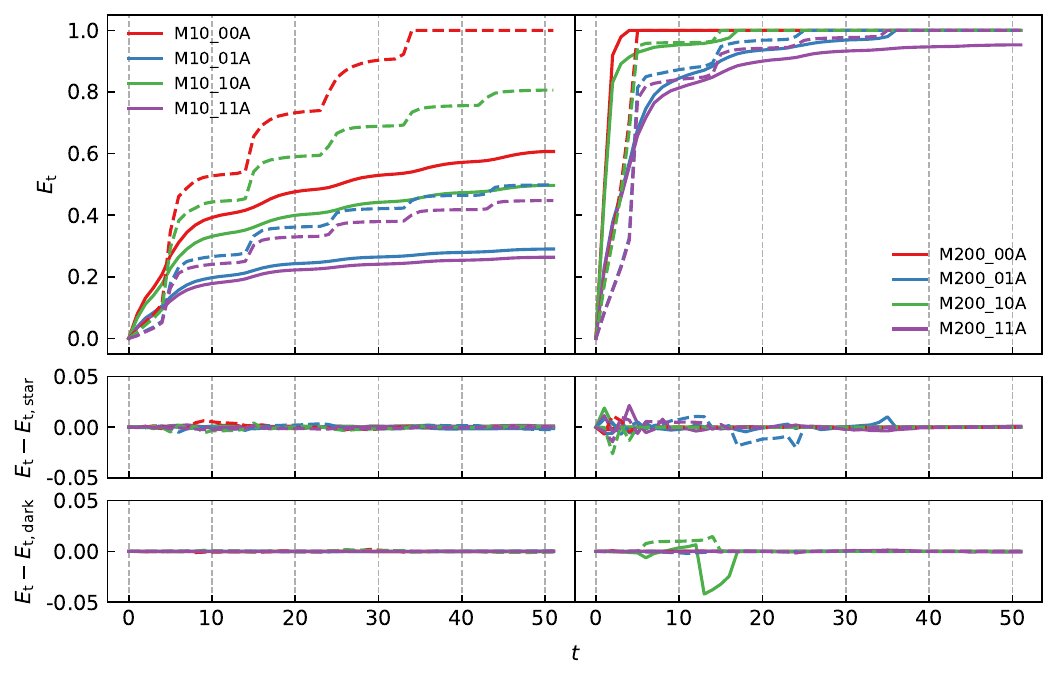}
\centering
    \caption{The plots show the energy truncation $E_\mathrm{t}$, as a function of $t_\mathrm{unit}$ and the component-wise residuals, $E_\mathrm{t} - E_\mathrm{t,star}$ and $E_\mathrm{t} - E_\mathrm{t,dark}$. The left column of plots corresponds to M10 simulations, and the right column shows M200 simulations. The colors indicate the $\gamma$-pair combination with solid lines showing orbit A and dashed lines showing orbit B. The plots show that energy truncation for any $\gamma$-pair is the same for both the star and dark components and that deviations for this energy truncation from $E_\mathrm{t}$ are minimal.}
    \label{fig:truncation}
\end{figure*}

We calculate the truncation energy $E_\mathrm{t}$
as the energy at which 50\% of the initial (dark matter and stellar) particles remain. This is tracked across all snapshots for each simulation, as shown in Figure~\ref{fig:truncation}. The plot shows the truncation energy $E_t$ and the residuals $E_t-E_{t,\mathrm{comp}}$, where $E_{t,\mathrm{comp}}$ is the energy at which 50\% of particles in the component are lost, as the satellites evolve. The residuals confirm that the components are tidally stripped to the same $E_t$ value (with deviations typically less than 0.5 per cent), though the stellar component exhibits larger deviations (i.e., more noise) because there are fewer stellar particles than dark matter particles.

\begin{figure}
    \includegraphics[]{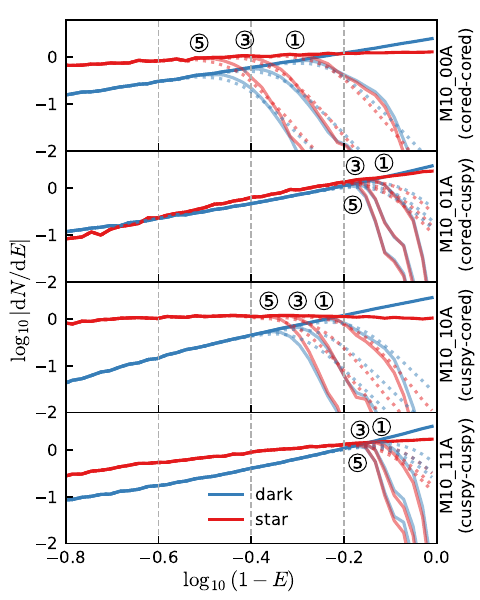}
    \caption{The energy distribution of particles of simulation M10\_XXA compared to the model from \cite{errani2022}. The circled numbers indicate the orbit $t/t_\text{orb}$ for the truncation lines. The y-axis is normalized by the total number of particles for that component. We find that for orbit A, the simulation has a sharper truncation compared to the \cite{errani2022} model.}
    \label{fig:errani_model}
\end{figure}

Finally, in Fig.~\ref{fig:errani_model}, we compare our results to the empirical model presented in \cite{errani2022, errani2024}, where each component evolves according to:
\begin{equation}
{\rm d}N/{\rm d} E^* \big|_{t} = \dfrac{{\rm d}N/{\rm d} E^* \big|_{t=0}}{1 + [0.85~E^*/E_{\rm mx}(t)]^{12}} \,\,\, 
\end{equation}
where 
\begin{equation}
E^* \equiv 1-E \,\,\, .
\end{equation}
We calculate $E_{\rm mx}$ by finding the energy where 99 percent of the original particles remain bound.\footnote{\citep{errani2022} also provides a fitting formula for this quantity, but it is sensitive to the initial profile used.}

We find that the model reproduces the overall evolution of the energy distribution reasonably well, and predicts stripping to the same energy level in the stellar and baryonic components, consistent with our findings. On the other hand, the steepness of the truncation in our simulations depends on orbital properties. For the A orbit, the simulated truncation is noticeably sharper than predicted by the empirical model. This is expected because the A orbit is relatively circular, producing a well-defined tidal boundary in energy space, as discussed above. In contrast, the B orbit (not shown) exhibits a much more gradual truncation than the \cite{errani2022, errani2024} model. These results suggest that the empirical model captures the behavior of a typical orbit, but extending the model to include an explicit dependence on orbital properties may provide a more accurate description of tidal evolution across a broader range of systems.

\section{A case study: NGC 1052-DF2} \label{sec:real_gal}

To relate our results to a more concrete example, we consider the possible evolutionary history of NGC 1052-DF2 (hereafter DF2). DF2 is a nearby ultra-diffuse galaxy (UDG), located approximately 20 Mpc away, characterized by its low surface brightness, extended morphology, and unusually low velocity dispersion among its globular clusters \citep[GCs; ][]{vandokkum2018}. Based on dynamical mass estimates derived from the GCs, \citet{vandokkum2018} argued that DF2’s total mass may be consistent with its stellar mass alone, suggesting little to no dark matter content---a result that challenges the conventional view that all galaxies form within massive dark matter halos. 

Many formation scenarios have been proposed to explain the unusual properties of DF2, and another similar dwarf galaxy in the NGC 1052 group, DF4, including tidal stripping \citep[e.g.,][]{ogiya2018, carleton2019, nusser2020, ogiya2022, montero2024}, interaction with galaxies \citep[e.g.,][]{fensch2019, moreno2022}, feedback \citep[e.g.,][]{trujillo2021}  or, more recently, a high-speed “bullet-dwarf” collision \citep[e.g.,][]{lee2024,tang2025}. Additional observational support for a tidal stripping origin in DMD galaxies comes from NGC 1052-DF4, which exhibits tidal features consistent with ongoing disruption \citep{montes2020}. By contrast, similar tidal tails have not been detected around DF2 \citep{montes2021, golini2024}.

Here we explore the tidal stripping scenario for creating DF2, in which a galaxy with a cored dark matter halo loses most of its dark matter due to the tidal field of a larger host galaxy, leaving behind a stellar-dominated remnant. Simulations by \citet{ogiya2018,ogiya2022} have shown that this DF2 formation scenario is feasible under specific orbital and structural conditions. To study the tidal stripping scenario for DF2, we loosely follow the parameters and modeling choices outlined in \citet{ogiya2022}.

\begin{figure}[htp!]
\includegraphics[width=\columnwidth]{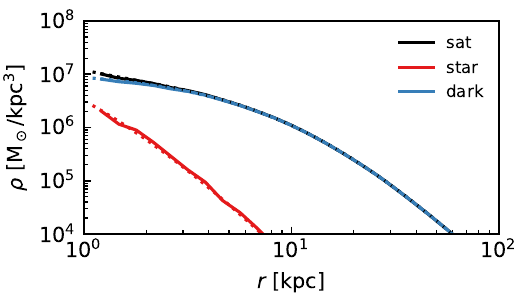}
    \caption{Initial conditions (dotted lines) of the DF2 galaxy, evolved in isolation over $t=5000\,t_{\rm unit}$ (solid lines). Red, blue, and black lines represent the stellar, dark matter, and total profiles, respectively.}
    \label{fig:NGC-stability}
\end{figure}

We model DF2 using a double-gamma profile. The cuspy stellar component has an inner slope of $\gamma_{\rm star}=0.44$, a scale radius of $r_{\rm s,star} = 1.1$ kpc and a total mass of $2 \times 10^8\, M_{\odot}$. The cored dark matter component has an inner slope of $\gamma_{\rm dark}=0$, a scale radius of $r_{\rm s,dark}=12.23$ kpc and a total mass of $9.15 \times 10^{10}\, M_{\odot}$ . 
The final DF2 model, and the stability of the initial conditions are shown in Figure~\ref{fig:NGC-stability}.

We evolve the dwarf galaxy in an NFW profile with $M_{\rm vir}=1.1\times10^{13} M_\odot$ in a  $r_{\rm vir}=458.76$ kpc, and $r_s=67.46$ kpc. 
We use an infall radius and velocity of 458.73 kpc and 61.85 km/s. This corresponds to circularity $\epsilon_{\rm circ}=0.3$, a relative energy\footnote{Defined as the orbital energy divided by the energy of a circular orbit at the virial radius.} of $x_c=1.39$, and an orbital period of $t_{\rm orb}=2.02 \times 10^8$ years.

\begin{figure}[htp!]
\includegraphics[width=\columnwidth]{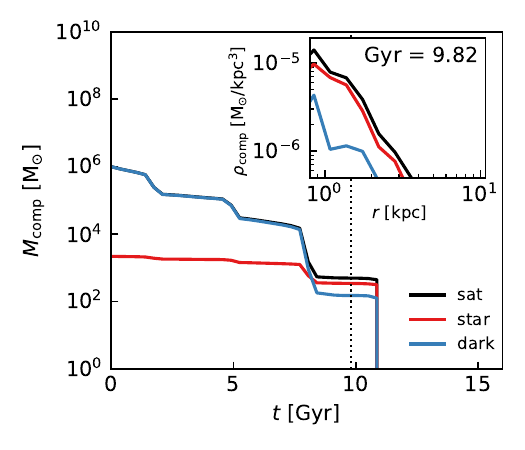}
    \caption{Mass loss of DF2 as a function of time since infall. The inset plot shows the satellite's density profile at $\text{Gyr}=9.82$ indicated by the dashed line on the main panel). DF2 is expected to have been accreted approximately 9 Gyr ago, and has since undergone significant tidal stripping. The stellar component remains bound longer than the dark matter, consistent with DF2's classification as a DMDG.}
     \label{fig:NGC-massloss}
\end{figure}

The resulting mass loss is shown in Figure~\ref{fig:NGC-massloss}. The dwarf galaxy becomes a DMD after its third pericentric passage and is disrupted thereafter. DF2 is expected to have been accreted $\sim 9$ Gyr ago \citep{ogiya2022}, which agrees with our plot. The inset plot shows the density profile at 9.82~Gyr in more detail. From these results, we expect that DF2 will be a DMD galaxy for one orbital period (approximately 200 Myr) before becoming disrupted.

\begin{figure}[htp!]
\includegraphics[]{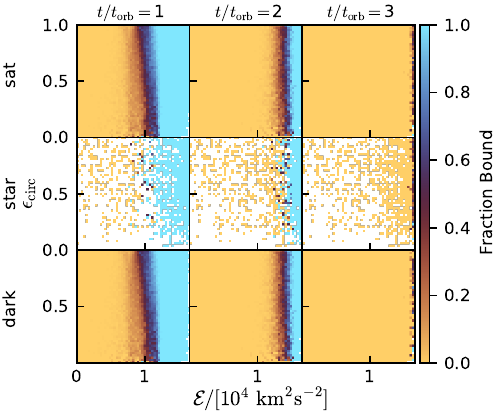}
    \caption{Stripping of DF2 satellite components based on initial circularity $L_0/L_{\rm max}$ and initial energy $\mathcal{E}_{\rm initial}/[10^4\ \text{km}^2\text{s}^{-2}]$. The colors indicate the percentage of particles that remain in the satellite. The top row shows the satellite total, the middle row shows the star component, and the bottom row shows the dark component. The columns show the orbits $t_{\rm orb}=1, 2, 3$ respectively. As before, both components are stripped to the same truncation energy.}
    \label{fig:DMDG-energy}
\end{figure}

As in the previous section, we find that both components of the satellite are stripped to the same energy, as shown in Figure~\ref{fig:DMDG-energy}. The remaining stars correspond to the most tightly bound material at infall, retaining memory of the original structure. For DF2, we estimate this truncation or ``tidal energy'' to be $1.6 \times 10^4\, {\rm km}^2\, {\rm s}^{-2}$. While this quantity is not directly measurable for DF2, it could potentially be inferred from the velocity dispersion of tidal streams, as discussed in Section~\ref{sec:discussion}.

\section{Discussion} \label{sec:discussion}

This study demonstrates that both stellar and dark matter components in multi-component systems are stripped to the same truncation energy, regardless of differences in their inner density slopes, $\gamma$. The behavior of tidal stripping in energy space provides a physically motivated framework for understanding the structural diversity of dwarf galaxies, including the emergence of DMD systems. Our results support the use of energy-based approaches to tidal stripping \citep[as developed in, e.g.,][]{drakos2017, drakos2020, drakos2022, stucker2021, stucker2023}, as a natural tool for modeling satellite evolution.

In our suite of simulations, we find that systems with a cored dark matter profile and a cuspy star profile will always evolve towards a system that is mainly stellar matter. This supports the hypothesis that tidal stripping can lead to DMD systems. The timescale that these DMDGs survive before becoming fully disrupted depends on the specific orbital parameters, but typically, the remnant survives for one orbital period before becoming disrupted at its next pericentric passage. 

A central question in tidal stripping studies is whether cuspy systems can ever be fully disrupted. While artificial disruption is known to be common in simulations \citep{vandenbosch2018, vandenbosch2018b, benson2022}, idealized studies generally find that cuspy subhalos are resilient and can survive indefinitely \citep[e.g.,][]{errani2020, amorisco2021, drakos2022}. However, it remains uncertain whether this robustness persists in more complex, realistic environments \citep[e.g.,][]{he2024, riley2024, shipp2024}. If central density is conserved, this has important implications for dark-matter annihilation signals \citep{drakos2023}. Our findings suggest that if either the stellar or dark matter component is cuspy, it provides a form of ``protection" to the other component, reducing the likelihood of complete disruption. However, additional work is required to confirm whether this protective effect is robust across a broader range of initial conditions and environments.

Our results suggest that tidal stripping operates most simply in energy space, stripping both stellar and dark matter components at the same binding energy, regardless of their detailed density profiles.
A tantalizing implication of this result is that the energy of observable stripped stars, such as those in stellar streams, could allow us to infer key characteristics of the dark matter structure and orbital history.
This connection opens up a new approach for constraining the dark matter content of disrupted galaxies by linking observable stellar kinematics to the energy-space stripping framework. 

Specifically, if stars and dark matter are stripped at the same truncation energy $\mathcal{E}_t$, then the measured stellar velocity dispersion provides an estimate of $\mathcal{E}_t$, which in turn depends on the bound dark matter mass and tidal radius. Combined with an independent estimate of the orbital tidal field, this allows the bound dark matter mass $M_{\rm dark}$ to be inferred from observables alone. In any self-bound system, the stellar velocities must remain below the escape speed, offering a lower bound on $\mathcal{E}_t$ and hence on $M_{\rm dark}$. In future work, we will investigate how reliably $\mathcal{E}_t$ can be estimated from observed velocity distributions and how variations in orbital parameters affect the robustness of this method.

Finally, it is important to acknowledge the limitations of our approach. In reality, galaxies are influenced not only by tidal forces but also by complex baryonic processes such as stellar feedback, gas dynamics, and cosmic rays. For example, work by \cite{koudmani2025} demonstrates that feedback mechanisms, including AGN activity and cosmic rays, can significantly alter dark-matter density profiles in dwarf galaxies, creating a spectrum of outcomes from cuspy to cored structures. These processes are not captured in our collisionless simulations, but likely play a critical role in shaping the structural evolution of subhalos. Incorporating such baryonic effects into future studies will be essential for a more complete understanding of the interplay between tidal forces and internal feedback, particularly in the context of explaining the observed diversity of rotation curves in dwarf galaxies.

In conclusion, this study provides a framework for understanding tidal stripping in multi-component systems, offering new insights into the structural diversity of galaxies. By linking the energy distribution of stripped stellar components to the properties of their dark matter halos and orbital histories, our results highlight the predictive power of energy-based models in explaining the evolutionary pathways of dwarf galaxies.

\section*{Acknowledgments}

This work was catalyzed during the Lamat REU program supported by NSF grant 1852393. The authors are grateful to the Lamat program for creating an engaging and productive research environment. J.~E.~T. acknowledges support from the Natural Sciences and Engineering Research Council of Canada (NSERC), through a Discovery Grant. N.~E.~D acknowledges support from NSF grants LEAPS-2532703  and AST-2510993. The authors would also like to thank Aaron J.~Romanowsky, Yimeng Tang, Raphaël Errani, and the anonymous referee for useful discussions.

The simulations for this study were performed on computing clusters from the lux supercomputer at UC Santa Cruz, funded by the National Science Foundation MRI Grant No.~AST 1828315.

\emph{Software}: \textsc{Gadget-2} \citep[]{springel2005}, \textsc{ICICLE} \citep[]{drakos2017}, \textsc{Numpy} \citep[]{harris2020}, \textsc{Matplotlib} \cite[]{hunter2007} and \textsc{Scipy} \citep[]{virtanen2020}.

\bibliographystyle{aasjournal}
\bibliography{two_component}

\end{document}